\documentclass[aps,prl,twocolumn,groupedaddress]{revtex4-1}
\usepackage{graphicx}
\usepackage{caption}
\DeclareCaptionLabelSeparator{dot}{. }
\makeatletter
\def\justified{
	\let\\\@normalcr
	\@rightskip\z@skip \rightskip\@rightskip
	\leftskip\z@skip
	\parindent 0em\relax
	\setlength{\parfillskip}{0pt plus 1fil}}
\DeclareCaptionJustification{justified}{\justified}
\usepackage{subcaption}
\usepackage{amsmath}
\usepackage{braket}
\usepackage{units}
\usepackage{ragged2e}
\usepackage[colorlinks,urlcolor=blue ,citecolor=blue ,linkcolor=blue ]{hyperref}
\usepackage{xcolor}
\usepackage{nicefrac} 
\usepackage{lipsum}
\usepackage{nameref}
\usepackage{hyperref}
\usepackage{textcomp} 

\newcommand{\bs}{\boldsymbol}
\newcommand{\Er}{\ensuremath{^{166}}{\rm Er} }
\newcommand{\Eexc}{\ensuremath{\varepsilon}}

\newcommand{\as}{\ensuremath{a_{\rm s}}}
\newcommand{\add}{\ensuremath{a_{\rm dd}}}
\newcommand{\edd}{\ensuremath{\epsilon_{\rm dd}}}
\newcommand{\tho}{t_{\rm h}}
\newcommand{\tr}{t_{\rm r}}

\newcommand{\vq}{{\bs q}}
\newcommand{\krot}{q_{\rm rot}}

\newcommand{\asf}{\as}
\newcommand{\asth}{a_{\rm s}^{\rm th}}

\newcommand{\rvec}{\mathbf r}

\newcommand{\kBragg}{q}
\newcommand{\ThetaBragg}{\theta}
\newcommand{\lambdaBragg}{\lambda_L}
\newcommand{\tauBragg}{\tau}
\newcommand{\kySquared}{\langle q_y^2\rangle}
\newcommand{\kySquaredBar}{\langle q_y^2\rangle}
\newcommand{\ascritExp}{a_\textrm{s}^{*}}

\newcommand{\kBraggRot}{q_{\textrm{rot}}}
\newcommand{\DSFbrut}{S_0(\vq,\,\omega)}
\newcommand{\DSF}{\widetilde{S}_0(\vq,\,\omega)}
\newcommand{\DSFres}{\widetilde{S}_0(\kBraggRot,\,\omegaResRot)}
\newcommand{\omegaBragg}{\omega}
\newcommand{\DSFBragg}{\widetilde{S}_0(\kBragg,\,\omegaBragg)}

\newcommand{\NBEC}{N}
\newcommand{\Nexc}{N_\textrm{exc}}
\newcommand{\omegaRes}{\omega_\kBragg}
\newcommand{\fexc}{\mathcal{F}}
\newcommand{\fexcres}{\mathcal{F}_{\rm res}}
\newcommand{\omegaResk}{\omega_\kBragg}
\newcommand{\omegaResRot}{\omega_\textrm{rot}}
\newcommand{\omegaResMax}{\omega_\textrm{m}}

\newcommand{\V}{V_0}

\begin{document}
	
	\title{
		Probing the roton excitation spectrum of a stable dipolar Bose gas
	}
	\author{D.\,Petter,$^{1}$ G.\,Natale,$^{1}$ R.\,M.\,W.\,van\,Bijnen,$^{2}$ A.\,Patscheider,$^{1}$ M.\,J.\,Mark,$^{1,2}$ L.\,Chomaz,$^{1}$ and F.\,Ferlaino$^{1,2,*}$}
	
	\affiliation{%
		$^{1}$Institut f\"ur Experimentalphysik, Universit\"at Innsbruck, Technikerstra{\ss}e 25, 6020 Innsbruck, Austria\\
		$^{2}$Institut f\"ur Quantenoptik und Quanteninformation, \"Osterreichische Akademie der Wissenschaften, Technikerstra{\ss}e 21a, 6020 Innsbruck, Austria\\
	}
	\date{\today}
	
	\begin{abstract}
		We measure the excitation spectrum of a stable dipolar Bose--Einstein condensate over a wide momentum-range via Bragg spectroscopy. We precisely control the relative strength, $\edd$, of the dipolar to the contact interactions and observe that the spectrum increasingly deviates from the linear phononic behavior for increasing $\edd$. Reaching the dipolar dominated regime $\edd>1$, we observe the emergence of a roton minimum in the spectrum and its softening towards instability. We characterize how the excitation energy and the strength of the density-density correlations at the roton momentum vary with $\edd$. Our findings are in excellent agreement with numerical calculations based on mean-field Bogoliubov theory.
		When including beyond-mean-field corrections, in the form of a Lee-Huang-Yang potential, we observe a quantitative deviation from the experiment, questioning the validity of such a description in the roton regime.
	\end{abstract}
	
	\maketitle

The spectrum of elementary excitations is a key concept providing insight into the quantum behavior of many-body systems. An emblematic example is the one of superfluid helium. At low momentum, the interactions among particles lead to collective-excitation modes with a linear energy\,($\Eexc$)-momentum\,($q$) dependence. Those are known as phonons, highlighting their analogy to sound waves. In addition, the strong interactions in He induce pronounced correlations at the mean interparticle distance, $d$. Such correlations reveal themselves in an energy minimum in the excitation spectrum at $q \approx 1/d$, termed {\em roton}~\cite{Landau41}.
Its physical interpretation, and even its mere existence, has been intensively debated for decades~\cite{Griffin:1993}. 
In today's understanding, the roton relates to the system's tendency to establish a crystalline order~\cite{Nozieres2004itr}, possibly providing access to supersolid phases~\cite{Boninsegni:2012}. Between the phonon and the roton, a local energy maximum, termed maxon, appears.

For gaseous Bose--Einstein condensates (BECs), the excitation spectrum also embeds the many-body interacting behavior. 
In the weakly-interacting bulk regime, the excitation spectrum is well described within the Bogoliubov theory and takes the form 
$\Eexc(\vq)=\sqrt{E(q)^2+2E(q)V_{\rm int}(\vq)}$, with $E(q)\propto q^2$ being the free-particle energy and $V_{\rm int}(\vq)$ being the mean-field interaction energy contribution~\cite{Pitaevskii:2016}.
In the case of short-range (contact) interactions, $V_{\rm int}$ is independent of $\vq$ 
and a roton minimum is absent, as confirmed in experiments~\cite{Stenger:1999,StamperKurn1999eop,Steinhauer:2002,Ozeri:2005,Gotlibovych:2014}.  Deviations from the Bogoliubov theory were observed in the strongly interacting regime~\cite{Papp:2008,Lopes:2017}, yet a roton minimum has remained elusive~\cite{Chevy:2016}.

Quantum gases with dipole-dipole interactions (DDIs), underlying
a ${\bs q}$-dependence of  $V_{\rm int}$, bring a paradigm shift in the many-body behavior~\cite{Baranov:2008,Lahaye:2008,Kadau:2016,Chomaz:2016,Schmitt:2016}. In particular, dipolar BECs (dBECs) are predicted to support a roton mode in their Bogoliubov spectrum~\cite{Santos:2003,ODell2003}. This roton spectrum requires specific conditions, namely: (i) an anisotropic geometry, tighter along the dipole direction, and (ii) a dominant DDI over the contact interaction. These conditions enable $V_{\rm int}$ to depend and change sign with $q=|{\bs q}|$, yielding a local minimum in $\Eexc(q)$ for ${\bs q}$ along the weak confinement axes. Conditions (i-ii) also dictate the roton mode's characteristics: {its momentum, $\krot$,} is governed by the confinement length along the dipoles (i), and $\Eexc(\krot)$ is controlled by the ratio $\edd=\add/\as$ of the dipolar ($\add$) and $s$-wave  scattering ($\as$) lengths (ii). In particular, $\Eexc(\krot)$ decreases {(softens)} for increasing $\edd$ and ultimately vanishes, yielding a mean-field instability.
The existence of dipolar rotons has been demonstrated in recent quench experiments, via the exponential growth of the roton {mode's} population when $\Eexc(\krot)$ turns imaginary, i.\,e., in the roton instability regime~\cite{Chomaz:2018}. 

In this Letter, we directly probe the phonon-maxon-roton excitation spectrum of a stable dBEC of ultracold erbium atoms. By precisely controlling $\edd$ (via $\as$), we observe the emergence of a roton minimum at large momentum and study in detail its softening. Our spectroscopic approach is based on the well-established technique of Bragg spectroscopy~\cite{Stenger:1999,StamperKurn1999eop,Steinhauer:2002,Brunello:2001,Blakie:2002,Ozeri:2005,Gotlibovych:2014,Papp:2008,Lopes:2017,Blakie:2012,Yang:2018,Veeravalli:2008}. For dBECs, this technique has been previously applied on Cr in the regime of weak DDI~\cite{Bismut:2012}, proving the anisotropy of $\Eexc(\vq)$, and consequently of the speed of sound, recently confirmed with a different technique with Dy~\cite{Wenzel:2018}. Bragg spectroscopy has also been employed to observe roton-like minima in the dispersion relations of hybrid systems of short-range interacting atoms and light~\cite{Ji2015, LiChung15, Mottl12}.

\begin{figure}[t!]
	\includegraphics[width=1\linewidth]{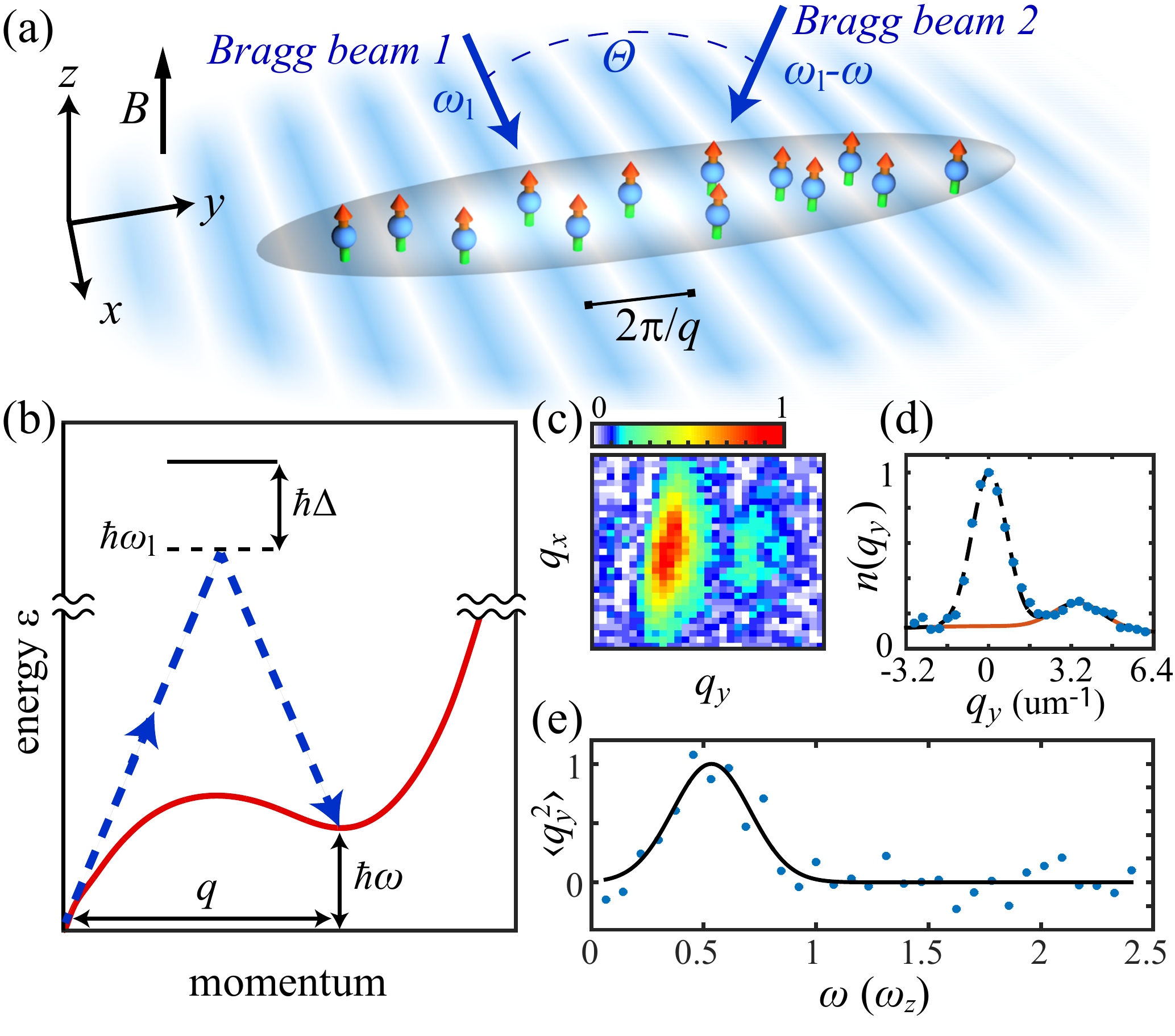}%
	\caption{\label{fig:setup}
		(a) The dBEC (gray ellipsoid) is axially elongated along $y$ and the atomic dipoles point along $z$. Two Bragg beams of frequency $\omega_{\rm l}$ and $\omega_{\rm l}-\omegaBragg$ (blue arrows) form a traveling grating along $y$ with a wavevector $\kBragg$ and velocity $\omegaBragg/\kBragg$ (blue shading). (b) The Bragg excitation drives a stimulated two-photon transition (dashed arrows), transferring a momentum $\kBragg$ and an energy $\hbar\omegaBragg$ to the atoms, resonant for $\hbar\omegaBragg = \varepsilon(q)$ (solid line). $\Delta\gg\,\omegaBragg$ is the detuning from the intermediate state.
		(c) Example of $n(q_x,q_y)$ after a Bragg excitation with ($\kBragg,\,\omegaBragg$)=(1.74(9)\,$l_z^{-1}$, $2\pi\times180$\,Hz). (d) Corresponding $n(q_y)$  (dots), fitted with a multi-Gauss function (dashed line). The solid line shows the component of the fit corresponding to the Bragg-excited atoms. (e) $\kySquaredBar$ vs. $\omega$ for $\kBragg=0.74(3)\,l_z^{-1}$. The solid line shows a Gaussian fit used for extracting $\omegaRes$ and normalizing the data.
	}
\end{figure}

Our Bragg spectroscopy is performed using a dBEC of strongly magnetic \Er atoms, prepared as in Ref.\,\cite{Chomaz:2016,Chomaz:2018}. After preparation, we confine the dBEC in a cigar-shaped optical dipole trap with harmonic frequencies $\omega_{x,y,z}=2\pi\times(261,27,256)\,$Hz. A homogeneous magnetic field, $B$, maintains spin-polarization of the sample in the lowest Zeeman sublevel, with atomic dipoles aligned along $z$; see Fig.\,\ref{fig:setup}\,(a). It also sets the value of $a_s$ via a magnetic Feshbach resonance (FR)~\cite{Chin2010fri}, centered at about $0$\,G, for which the $B-$to$-\as$ conversion has been precisely extracted with a $\pm 2\,a_0$-wide prediction interval in the $\as$ range here explored\,\cite{Chomaz:2016, Chomaz:2018}. Systematic uncertainties on $\as$ are estimated to be up to $\pm3\,a_0$~\cite{supmat}. The dipolar length, $\add=\mu_0\mu^2m/12\pi\hbar^2=65.5\,a_0$, results from the atomic magnetic moment, $\mu$, and mass, $m$, of $\Er$.
$\mu_0$ is the vacuum permeability and $\hbar=h/2\pi$ the reduced Planck constant. After preparation, $\as$ equals $67\,a_0$, corresponding to $\edd\approx1$.
In this geometry, a roton mode is expected to emerge along the  axial ($y$) direction for $\edd>1$ and {softens} for increasing $\edd$. The roton minimum appears at a momentum $\krot \sim 1/l_z$, with $l_z=\sqrt{\nicefrac{\hbar}{m\omega_z}}\approx0.5\,$\textmu m.

To reach $\edd>1$, we decrease $\as$ to the desired value by ramping $B$ closer to the FR's pole. The ramping time, $\tr$, is chosen to be long enough to ensure adiabaticity with respect to the tight trapping frequencies ($\tr>\nicefrac{1}{\omega_{x,z}})$ but short enough to avoid too strong three-body collisional losses near the FR. For the highest $\edd$, we find an optimal trade-off for $\tr=15\,$ms, defining our fastest ramp. This ramp is not fully adiabatic with respect to the axial dynamics. We observe small-amplitude breathing and sloshing modes along $y$, which we account for in our spectroscopic measurements~\cite{supmat}. After ramping $B$, we hold the atoms for a time $\tho$, after which we perform Bragg spectroscopy to probe the excitation spectrum of our dBEC of $N$ atoms.

Our Bragg spectroscopy setup is illustrated in Fig.\,\ref{fig:setup}\,(a)  and detailed in Ref.\,\cite{supmat}. In brief, it uses two coherent laser beams of wavevector $k_L=2\pi/\lambdaBragg$, with $\lambdaBragg=401\,$nm, propagating in the $z-y$~plane and intersecting each other under an angle $\ThetaBragg$. At the cloud's position, the beams form a light grating along $y$ of potential depth $\V$ and wavevector $\kBragg=2k_L\,\mathrm{sin}(\ThetaBragg/2)$.
The two beams have a small frequency difference, $\omegaBragg$, causing the grating to travel at a velocity $\omega/\kBragg$. 
A key feature of our setup is the wide dynamical tunability of $\ThetaBragg$.
This is obtained by creating the Bragg beams using holographic gratings~\cite{Zupancic:2016,supmat}, generated with a digital micromirror device\,\cite{LiChung15}. By uploading different holograms, we can vary $\ThetaBragg$, and accordingly $\kBragg$ from 0 to 1.8$\,l_z^{-1}$. Moreover, by employing hologram sequences and changing their display rate, $\omegaBragg$ can be directly varied, up to $\sim2\pi\times 1$\,kHz. 
In the experiment, we illuminate the dBEC with a Bragg pulse of duration $\tauBragg$. The value of $\tauBragg=7\,$ms is chosen to be long enough to minimize Fourier broadening of the frequency spectrum, and yet short with respect to a quarter of the axial trap period~\cite{Blakie:2002,Steinhauer:2002,Yang:2018}.  
Immediately after the pulse, we switch off the trap and let the cloud expand for 30\,ms. We then image the atoms along $z$ via standard absorption-imaging, from which we extract the momentum distribution of the cloud, $n(q_x,q_y)$.

\begin{figure*}[t!]
	\includegraphics[width=\textwidth]{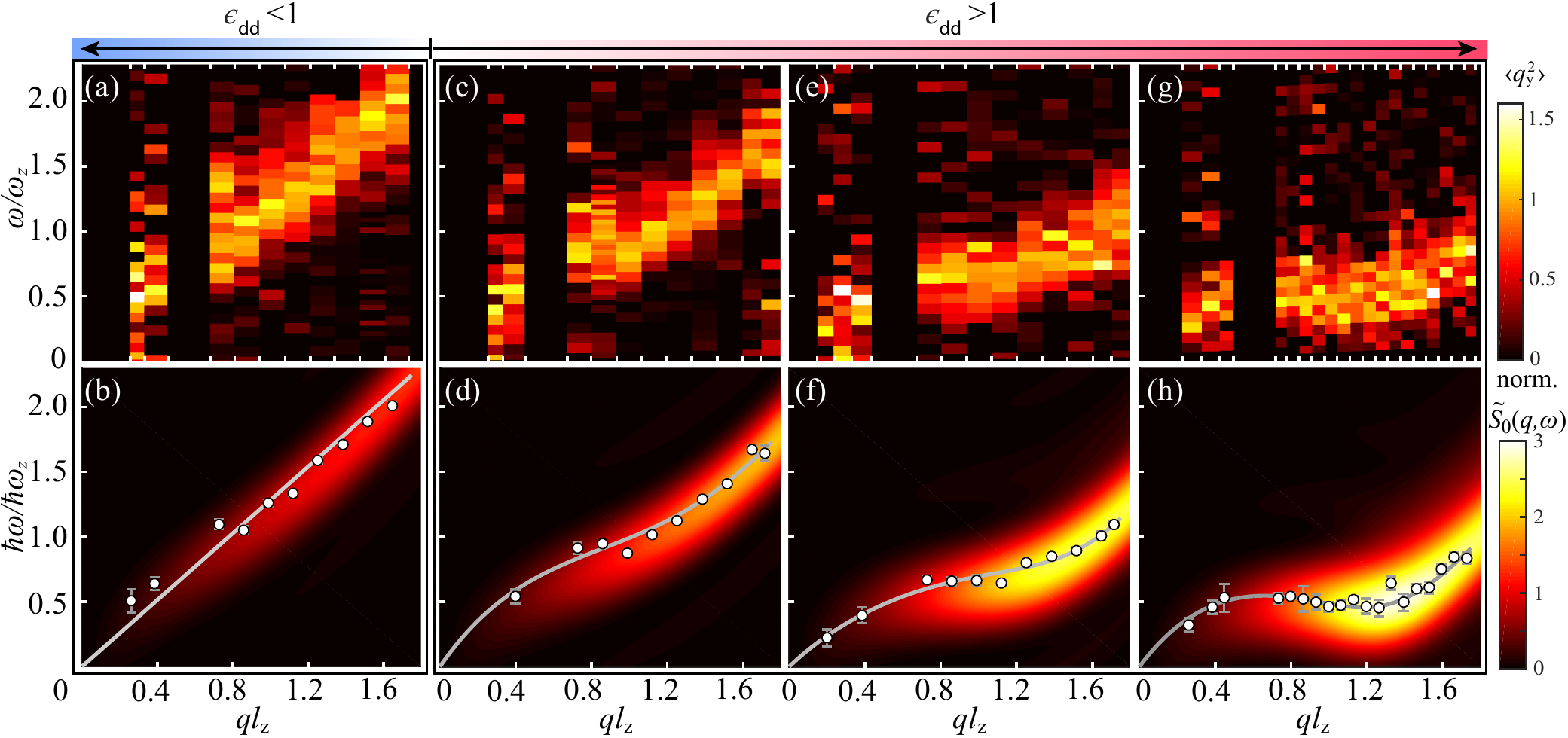}
	\caption{\label{fig:examples} 
		Excitation spectra from  $\edd<1$ to $\edd>1$: (a,\,c,\,e,\,g) Measured $\kySquared$ for varying $\kBragg$ (columns, delineated by white tick marks) and $\omegaBragg$ at given $\asf$.
		Each column is fitted with a Gaussian function and renormalized by the fitted peak amplitude. Black columns are inaccessible to measurements~\cite{supmat}.	(b,\,d,\,f,\,h) 
		Extracted $\Eexc(\kBragg)$ (white dots) from (a,\,c,\,e,\,g), respectively. Here and throughout the letter, the error bars denote $\pm$ one standard deviation.
		The solid lines are guides to the eye, based on the analytic formula from Ref.\,\cite{Blakie:2012}. The colormap shows the calculated $\DSFBragg$, normalized by the maximum of $\DSFBragg$ at $ql_z=1.3$ and $\as=82\,a_0$.
		For [(a,\,b);\,(c,\,d);\,(e,\,f);\,(g,\,h)], $\NBEC=[4.6(5);\,3.9(4);\,3.3(3);\,2.5(3)]\times10^4$ and $\as=(\as^{\textrm{exp}},\asth)=[(80.0,\,82.0);\,(60.5,\,62.5);\,(55.3,\,54.5);\,(52.5,\,51.6)]\,a_0$, respectively.
	}
\end{figure*}

The Bragg excitation can be interpreted as a stimulated two-photon transition, imparting a well-defined momentum, $\kBragg$, and energy, $\hbar\omegaBragg$, to the atoms;  see Fig.\,\ref{fig:setup}\,(b).
In bulk systems, for a fixed $\kBragg$ and varying $\omegaBragg$, atoms, initially at $q_y=0$, are resonantly transferred to $q_y=q$ for $\hbar\omegaBragg=\Eexc(\kBragg)$~\cite{Stenger:1999,StamperKurn1999eop,Steinhauer:2002,Ozeri:2005}. When accounting for finite-size effects, the response is broadened in $\kBragg$.
The dynamic structure factor, which quantifies the system's response to an external perturbation, can be related to the fraction of excited atoms during a Bragg pulse, $\fexc={\Nexc}/{(N_0+\Nexc)}$. Here, $N_0$ ($\Nexc$) is the number of the zero-momentum (Bragg-excited) atoms.
In the linear response regime~\cite{Brunello:2001,Blakie:2002,Ozeri:2005},  
\begin{equation}\label{eq:excitedFrac}
\fexc= {\frac{\pi^2\V^2\tauBragg}{h^2}\DSFBragg}
\end{equation}
where $\DSFBragg$ is the zero-temperature dynamic structure factor, Fourier-broadened in $\omegaBragg$ due to the finite $\tau$, see Ref.\,\cite{supmat}.

Figure~\ref{fig:setup}\,(c) shows a representative $n(q_x,q_y)$ for a $\kBragg\gtrsim\,l_z^{-1}$ excitation. For this high $\kBragg$, the Bragg-excited atoms are well resolved as a side peak. From a multi-Gauss fit to the integrated density, $n(q_y)$, we extract $\fexc$; see Fig.\,\ref{fig:setup}\,(d)~\cite{supmat}.
For $\kBragg\lesssim\,l_z^{-1}$, the zero-momentum peak and the Bragg-excited one overlap and $\fexc$ can not be precisely extracted. To access $\DSFBragg$ for all $\kBragg$, we use the momentum variance $\kySquared=\int n(q_y) q_y^2\,dq_y$, which relates to the imparted energy into the system. As $\fexc$, $\kySquared$ gives access to $\DSFBragg$, but via a more complex relation~\cite{Pitaevskii:2016, Brunello:2001,Blakie:2002,Meinert:2015,supmat}. Figure~\ref{fig:setup}\,(e)
exemplifies a resonance in $\kySquared$ when varying $\omegaBragg$ at fixed $\kBragg$. We extract its center frequency, $\omegaRes$, via a Gaussian fit.
By varying $\kBragg$ over the experimentally accessible range, we probe the lowest-lying branch of the
axial excitation spectrum $\Eexc(\kBragg)= \hbar\omegaRes$~\cite{supmat}.

Figure~\ref{fig:examples} shows the results of our Bragg measurements, revealing how $\Eexc(\kBragg)$ is modified when tuning from $\edd<1$ to $\edd>1$. For $\edd<1$, $\Eexc(\kBragg)$ shows a linear dependence over the whole $\kBragg$ range, characteristic of phonon modes; Fig.\,\ref{fig:examples}\,(a,\,b). From a linear fit to $\Eexc(\kBragg)$, we estimate the sound velocity $c=\lim_{q \rightarrow 0}{\nicefrac{\Eexc(\kBragg)}{\kBragg}}=1.01(1)\,$mm/s along $y$.
As we probe the system for increasing $\edd>1$, we find an overall reduction of the excitation energies and increasing deviations of the spectra from the linear phonon behavior; Fig.\,\ref{fig:examples}\,(c,\,d). When further increasing $\edd$, the spectrum starts to flatten at large $\kBragg$; Fig.\,\ref{fig:examples}\,(e,\,f). Ultimately, at the highest $\edd$, we observe a local minimum occurring at $\kBragg \approx \kBraggRot=1.27(6)\,l_z^{-1}$, providing an unambiguous signature of the existence of the roton mode; Fig.\,\ref{fig:examples}\,(g,\,h).  At intermediate momenta between the phonon and roton regimes, a maxon (local maximum in $\Eexc(\kBragg)$) is also identifiable. Due to optical constraints on our Bragg setup, the maxon regime is not fully accessible; see black region in Fig.\,\ref{fig:examples}\,(e,\,g).
To compare our measurements with theory, we perform calculations of $\DSFBragg$, by calculating the Bogoliubov modes from the Gross--Pitaevskii equation (GPE) linearised around equilibrium at the final $\as$ \cite{Ronen:2006,Blakie:2012,supmat}. Here we explicitly do not include beyond-mean-field effects~\cite{supmat}; see later discussion.
Over the entire range of $\edd$, our theory describes the experimental data, 
both qualitatively and quantitatively.
In the calculations of Fig.\,\ref{fig:examples}, we let $\as$ vary within the prediction interval ($\pm 2 a_0$) of our $B$-to-$a_s$ conversion to best match the measured spectrum.

\begin{figure}[]
	\includegraphics[width=1\linewidth]{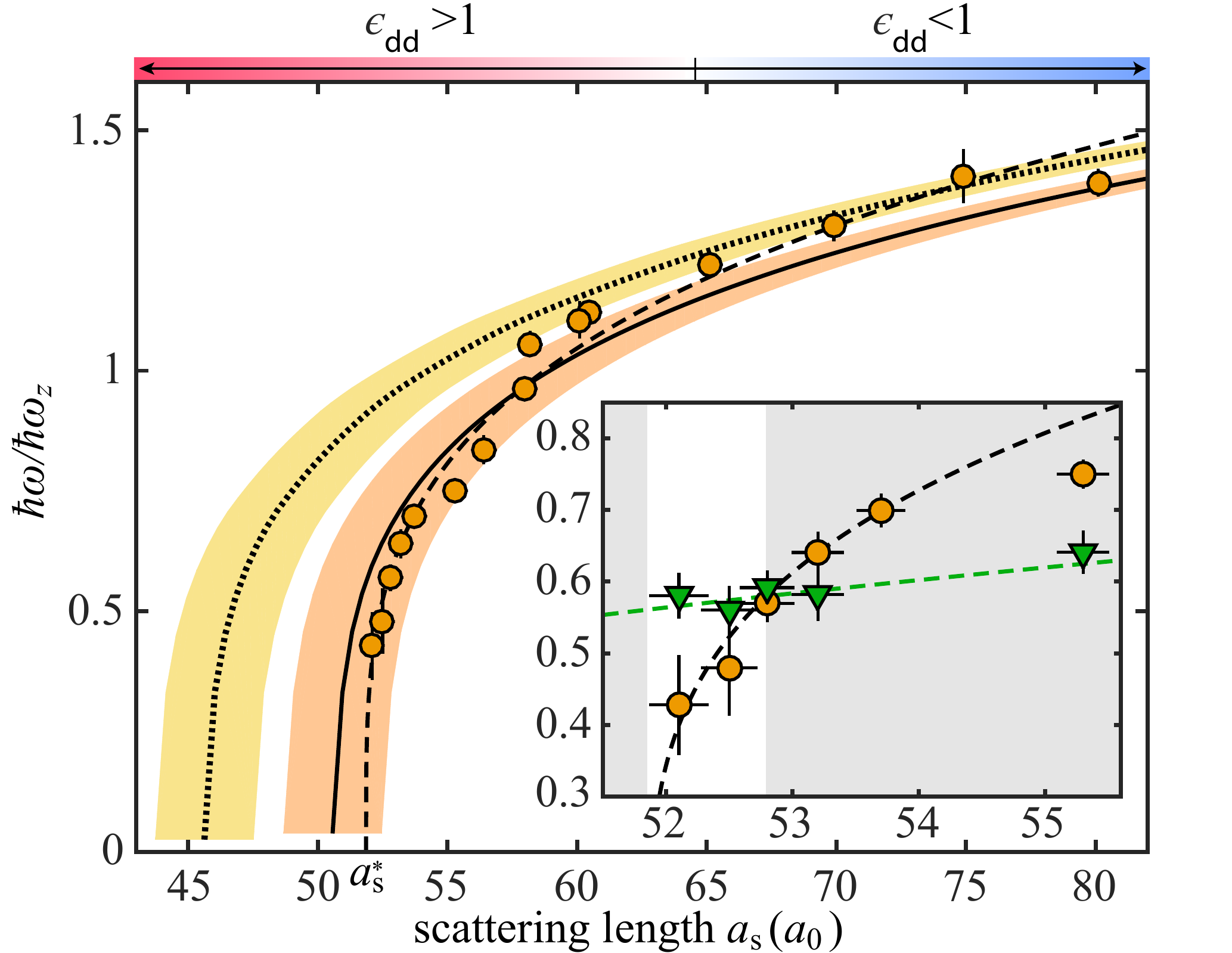}%
	\caption{\label{fig:energyScaling}
		Measured $\hbar\omegaResRot$ with $\NBEC=2.4(2)\times10^4$ vs.\,$\asf$ (circles) and corresponding numerical calculations excluding (solid line) and including (dotted line) beyond-mean-field effects. The shadings show the calculations over the prediction interval of $\as$. The dashed line shows the power-law fit to the experiment.
		Inset: zoom-in around $\ascritExp$ comparing $\hbar\omegaResRot$ (circles) to $\hbar\omegaResMax$ (triangles), respectively with the power-law fit and a guide to the eye (dashed lines). The $\as$-region where $0\leq\hbar\omegaResRot\leq\hbar\omegaResMax$ is highlighted with a white background.
		The error bars in $\hbar\omega_{\rm rot, m}$ ($\asf$) are the fit's statistical uncertainties (uncertainties from experimental $B$-field fluctuations~\cite{supmat}).
	}
\end{figure}
To get a deeper insight into the roton softening, we perform Bragg measurements at a fixed $\kBragg=\kBraggRot$ and extract $\omegaRes$, denoted $\omegaResRot$, as a function of $\asf$ for fixed $\NBEC$. As shown in Fig.\,\ref{fig:energyScaling}, $\omegaResRot$ exhibits a reduction that becomes increasingly sharp for decreasing $\asf$. Below $52\,a_0$, we observe that the system undergoes a roton instability, i.\,e.\,a spontaneous population of the roton mode even without applying a Bragg pulse; see also Ref.\,\cite{Chomaz:2018}. We find that the softening of $\omegaResRot$ is well approximated by an $\as$-power-law scaling. By fitting the data to $\omegaResRot(\as)=A(\as-\ascritExp)^{p}$, we extract the critical scattering length at which $\omegaResRot$ vanishes, $\ascritExp=51.9(2)\,a_0$, matching our instability observation.
We also observe a scaling exponent of $p=0.27(2)$ ($A$ is a scaling coefficient). 
The pronounced dependence of the roton energy on the interparticle interactions, i.\,e.\,on both $\as$ and the atomic density, makes the measurements at low energy very sensitive to fluctuations. Indeed, small fluctuations and drifts in $B$ and $N$ can already drive the system into instability, eventually preventing a reliable measurement of the spectrum for $\omegaResRot\lesssim 2\pi\times100\,$Hz; e.\,g.\,see horizontal error bar in the inset. 
For comparison, we additionally probe the $\as$-dependence of the excitation energy $\omegaRes$ near the maxon at $\kBragg=0.74(3)\,l_z^{-1}$, denoted $\omegaResMax$.
We observe that $\omegaResMax$ decreases much slower than the roton case. 
As shown in the inset of Fig.\,\ref{fig:energyScaling}, the two modes' energies cross around $\asf=52.8\,a_0$. For $\ascritExp<a_s<52.8\,a_0$, $\omegaResRot<\omegaResMax$, showing the emergence of a local minimum in the spectrum of a stable dBEC. At $a_s=52.2(2)\,a_0$, the minimum can be distinguished with a confidence level of 98\%~\cite{supmat}.

Figure~\ref{fig:energyScaling} also shows $\omegaResRot$ extracted from our numerical calculations, together with its variation within the prediction interval of $\as$. The theory describes our observation very well and confirms the rapid variation of the roton energy with $\as$.
We have also performed calculations including beyond-mean-field effects in the form of a Lee-Huang-Yang correction in the GPE~\cite{Lee1957eae,Lee1957mbp,Pelster:2011,Pelster:2012}. This additional term has proven to be crucial to understand the behavior of a dBEC in the droplet regime~\cite{Waechtler:2016,Bisset:2016,Waechtler:2016b,Chomaz:2016, Schmitt:2016}.
Interestingly, the agreement between theory and experiment becomes worse with a discrepancy that can not be accounted for with the experimental $\as$ uncertainty. Such a discrepancy can have several origins. These range from additional experimental uncertainties (e.\,g.\,$\NBEC$ values, effects of residual density-dependent dynamics~\cite{supmat}) to more fundamental reasons. As speculated in Ref.\,\cite{Chomaz:2018}, this mismatch could call into question the validity of standard treatments of beyond-mean-field effects in the roton regime. For instance, the standard inclusion of a Lee-Huang-Yang term in the GPE relies on a local density approximation and is justified for negligible quantum depletion and higher-order corrections~\cite{Pitaevskii:1998,Fabrocini:1999,Fabrocini:2001,Banerjee:2001,Waechtler:2016,Bisset:2016,Waechtler:2016b, Pelster:2011,Pelster:2012, Fu:2003, Astrakharchik:2018,Papp:2008,Giorgini:1998did}. These conditions might not be completely fulfilled in the roton regime. Future theoretical efforts, combined with stringent validity tests on experiments, are needed to shed light on this important aspect.

\begin{figure}[t!]
	\includegraphics[width=1\linewidth]{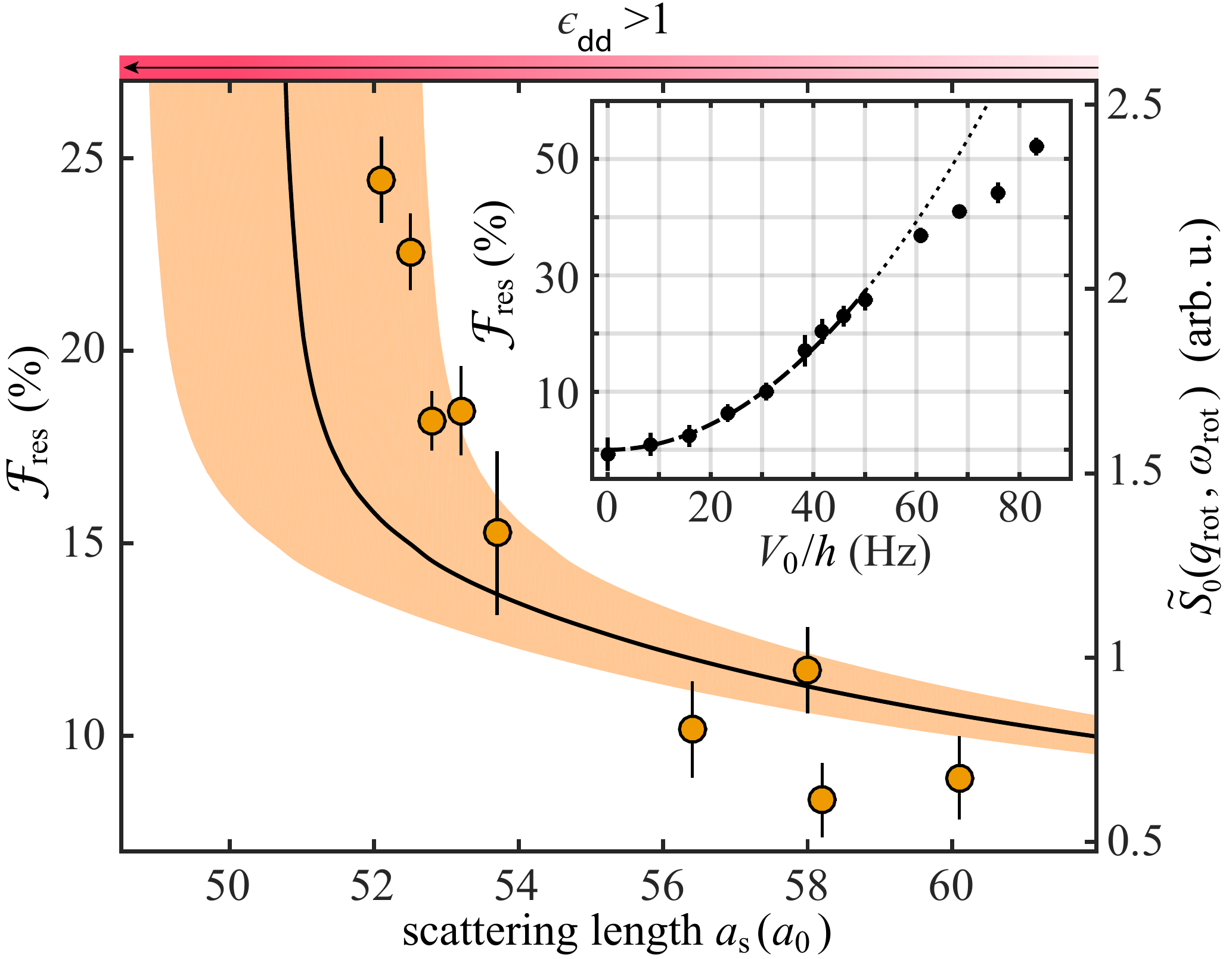}%
	\caption{\label{fig:Skrot}
		Measured $\fexcres$ with $\NBEC=2.4(2)\times 10^4$ and $\V/h=42$\,Hz (circles) and calculated $\DSFres$ (solid line) vs.\,$\asf$. The shading shows the theory over the prediction interval of $\as$.
		Inset, measured $\fexcres$ (circles) vs.\,$\V$ at $\asf=52.2\,a_0$. A quadratic fit up to $\V/h=50\,$Hz (dashed line) shows deviations to the data for higher $\V$ (dotted line).
	}
\end{figure}

The emergence of a roton minimum intrinsically connects to an increase of density-density correlations. This is quantified by the amplitude of $\DSFres$~\cite{StamperKurn1999eop,Blakie:2002,Blakie:2012,supmat}, which is related via Eq.\,\eqref{eq:excitedFrac} to the fraction of excited atoms at the Bragg resonance, $\fexcres$. In the experiment, we explore this aspect by measuring $\fexc$ as a function of $\omega$ at $\kBragg=\kBraggRot$ using a fixed $\V$. From a Gaussian fit to the data, we extract $\fexcres$. We repeat the experiment for various $\as$ in the $\edd>1$ regime; see Fig.\,\ref{fig:Skrot}. 
When approaching the roton instability ($\ascritExp$), $\fexc$ soars, with an increase by about a factor of 3 when changing $\as$ by less than 15\%. Such a behavior is also confirmed by our theoretical calculations. Experimentally, we find that the linear-response regime, i.\,e.\,the validity of Eq.\,\eqref{eq:excitedFrac}, extends up to $\fexcres\approx 25\%$; see inset of Fig.\,\ref{fig:Skrot}.

In conclusion, we measure the excitation spectrum of a dBEC and its evolution from the contact-dominated to the dipolar-dominated regime. In the latter regime, we observe	the emergence of a roton minimum. Comparisons with theory reveal a good agreement with mean-field Bogoliubov calculations and show deviations when including beyond-mean-field corrections, calling for further studies of their effects and their treatment in the roton regime. Similar to the cases of superfluid helium~\cite{Boninsegni:2012,Balibar:2010,Kirzhnits:1971cco,Schneider:1971} and of hybrid systems of atoms and light~\cite{Leonard:2017,Li:2017}, the roton minimum may provide a path for the creation of supersolid or crystalline phases in dBECs~\cite{Boninsegni:2012,Ancilotto:2018,Tanzi:2018,Wenzel:2018,Baillie:2018}. With the achievement of a precise knowledge and control of the roton softening, our work provides a first step in this direction.
	
	\begin{acknowledgments}
		We thank B.\,Blakie, S.\,Stringari, and L.\,Santos for inspiring discussions and fruitful comments on our manuscript. Additionally, we acknowledge D.\,Baillie, F.\,Dalfovo for stimulating discussions, S.\,Burchesky for help during initial offline tests, and S.\,Baier for the assistance during the early stage of the experiment.
		We acknowledge Swarovski Optik (S.\,Kroess and his team) for technical support on processing optics for the setup. Part of the computational results presented have been achieved using the HPC infrastructure LEO of the University of Innsbruck.
		This work is financially supported through an ERC Consolidator Grant (RARE, no.\,681432) and a DFG/FWF (FOR 2247/PI2790).
	\end{acknowledgments}
	
	* Correspondence and requests for materials
	should be addressed to F.\,F.~(email: francesca.ferlaino@uibk.ac.at).


	
	%

	\section*{Preparation and calibration of the \lowercase{d}BEC\lowercase{s}}
	\label{sec:trap}
	A dBEC of $\Er$ is prepared in the same way as described in  Refs.\,\cite{Chomaz:2016,Chomaz:2018}. Trapping is provided by crossed optical beams forming a harmonic potential $V(\rvec) = m (\omega_x^2 x^2 + \omega_y^2 y^2+ \omega_z^2 z^2)/2$ for the atoms. At the end of the preparation procedure, $V(\rvec)$ has a cigar-shaped geometry with $\omega_{x,y,z}=2\pi \times(261,27,256)\,$Hz. The frequencies are measured via exciting and probing either the center-of-mass oscillation of dBECs (for $\omega_x$ and $\omega_z$) or the breathing mode of cold, thermal samples (for $\omega_y$). The uncertainties of the trapping frequencies are at the few-percent level.
	After reshaping the trap, we ramp $\as$ linearly from  $\as =67\,a_0$ to its final value in a time $\tr$, by performing a corresponding ramp in $B$, computed from the calibrated $B-$to$-\as$ conversion~\cite{Chomaz:2016,Chomaz:2018}. The ramp time is chosen to be relatively long, $\tr\geq 15\,$ms; see main text. In our Bragg spectroscopy measurements, we apply the Bragg pulse after an additional holding time $\tho$.
	
	The number of atoms in the probed dBEC, $\NBEC$, is extracted from time-of-flight (TOF) measurements, performed using the same experimental sequence as for the Bragg measurements, but, instead of applying the Bragg pulse, simply waiting $\tho+\tau/2$ before releasing the atoms from the trap. We extract the integrated density distribution from standard absorption-imaging technique after 30\,ms of TOF. We fit a two-dimensional bimodal function made of a Gaussian and an inverted parabola at the power $3/2$ to the density distribution. The values of $\NBEC$ reported in the main text corresponds to the number of atoms in the parabolic peak. We note that $\NBEC$ typically fluctuates by up to 10\% between experimental runs. In addition, by measuring $\NBEC$ at $\tho$ and at $\tho+\tau$, we observe three-body losses during the Bragg pulse up to 20\% for the lowest values of $\asf$.
	Finally, we point out that the bimodal function employed to extract $\NBEC$ is an approximate description of a finite temperature BEC and may lead to an underestimation of $\NBEC$, especially at our lowest $\as$ values. The experimentally calibrated $\NBEC$ and $\omega_{x,y,z}$ are used as fixed parameters in the theory calculations; see below.
	
	\section*{Accounting for the breathing and sloshing modes: timing and Doppler shifts}
	\label{sec:Dopplershift}
	\begin{figure}[t!]
		\includegraphics[width=1\linewidth]{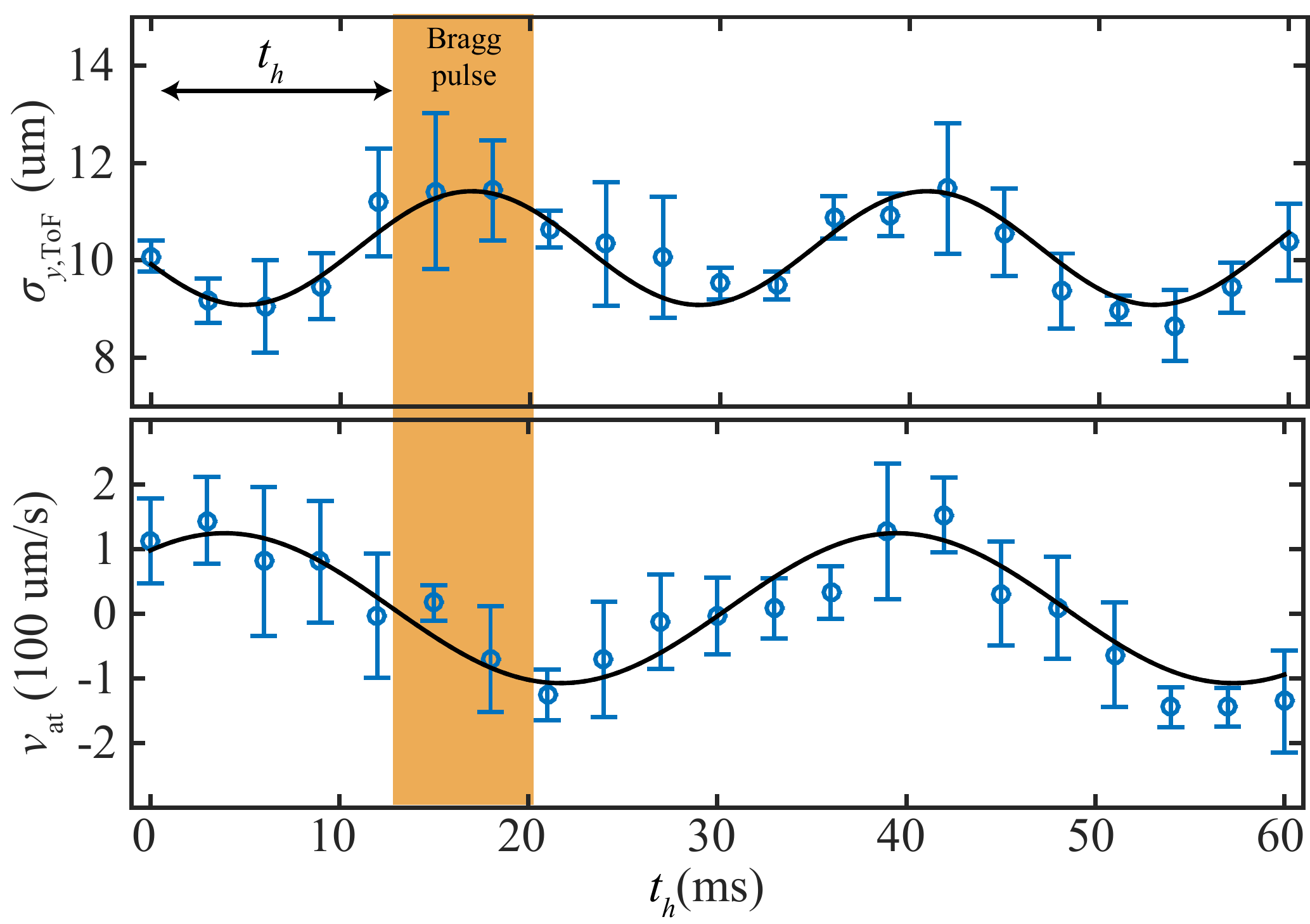}
		\caption{\label{figSupMat:CoM}
			Example of calibration measurements of the breathing and sloshing modes for the measurements at $\asf=52.5\,a_0$ of Fig.\,2\,(g,\,h) of the main text. We record the variations with $\tho$ of the cloud's size (upper panel) and position in TOF. From the position we compute the mean atomic velocity $v_{\rm at}(\tho)$ (lower panel). The solid lines are sinusiodal fits to the data. We time the Bragg pulse (orange shaded area) to be centered on the cloud's size maximum. The corresponding mean atomic velocity is in this case $v_{\rm at}=-68(16)$\textmu m/s. Similar measurements and analyses have been performed for each measurement reported in the main text.
		}
	\end{figure}
	Our fastest $\as$-ramps are not fully adiabatic with respect to the axial dynamics and may induce small-amplitude breathing and sloshing modes along $y$. Such excitations could affect our Bragg measurements. The former mode induces density oscillations and can influence the value of the roton excitation energy. The latter mode causes a sizable momentum-dependent Doppler shift of the Bragg excitation frequency~\cite{Stenger:1999}. We account for these effects in our experiment by performing dedicated calibration measurements. In particular, we probe the evolution of the atomic density distribution after TOF, as a function of $\tho$. We perform such calibration measurements for each $\as$-ramp employed in the experiment. 
	
	The breathing excitation reveals itself in the evolution of the axial size; see Fig.\,\ref{figSupMat:CoM} (upper panel). To minimize the impact of the breathing mode on our measurements, we synchronize the Bragg pulse symmetrically around the moment at which the size in TOF reached its maximum. Then, the in-situ density of the dBEC changes the least and remains close to its highest value during probing. The corresponding $\tho$, after which we switch the Bragg beams on, is typically between 10 and 20\,ms.
	
\begin{figure}[t!]
		\includegraphics[width=1\linewidth]{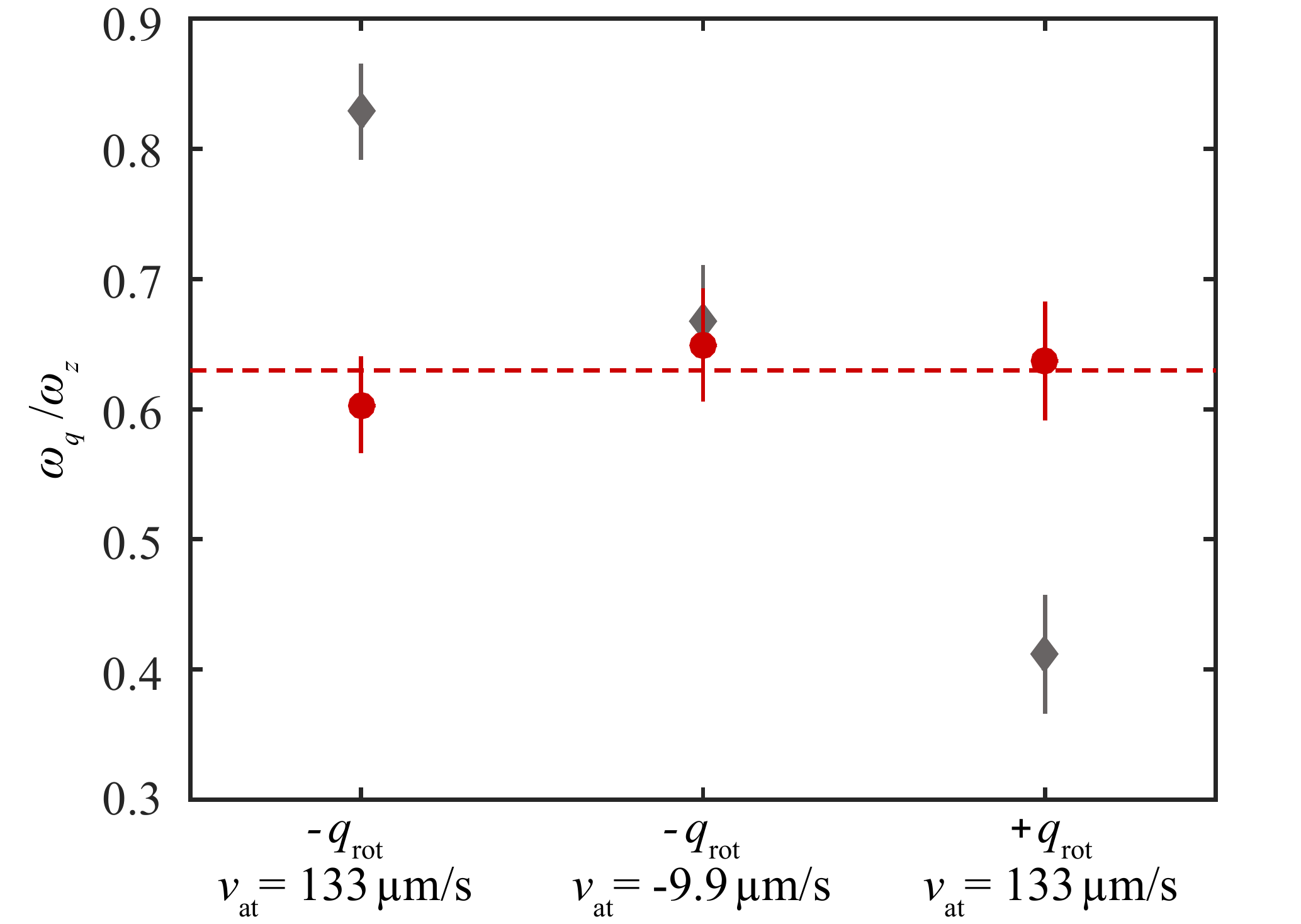}
		\caption{\label{figSupMat:DopplerCorrection}
			Bragg resonance frequencies $\omegaResk$ at $|\kBragg|=\kBraggRot$ and $\asf=52.7\,a_0$ measured using three different Bragg pulses, characterized by the couple  $\{\kBragg,v_{\rm at}\}$ (abscissa's labels). For each measurement, both the uncorrected (diamond) and Doppler-corrected (circle) frequencies are shown.	The dashed line indicates the mean of the Doppler-corrected resonance frequenies.
		}
	\end{figure}

	The sloshing mode reveals itself in the variation of center-of-mass position of the atomic cloud. This gives direct access to the mean velocity, $v_\textrm{at}(\tho)$, of the atoms in the dBEC as a function of $\tho$; see Fig.\,\ref{figSupMat:CoM} (lower panel). By averaging over the duration of the Bragg pulse, $v_\textrm{at}=\langle v_\textrm{at}(\tho) \rangle_{\tau}$, 
	we extract the induced Doppler shifts for the Bragg excitation, $\omega_\textrm{D}=v_\textrm{at}\kBragg$ which we then use to correct the applied Bragg frequencies $\omegaBragg$. To check the accuracy of our treatment, we have repeated Bragg spectrscopy measurements at various $\asf$ and $\kBragg$ using Bragg pulses corresponding to distinct $\omega_\textrm{D}$. In particular, to achieve distinct $\omega_\textrm{D}$, we reversed the Bragg excitation direction to compare measurements with $\pm\kBragg$ and used pulses starting at different $\tho$, yielding $v_\textrm{at}\approx \{-v_\textrm{max}, 0, v_\textrm{max}\}$, $v_\textrm{max}$ being the maximum insitu mean velocity. A set of such measurements is exemplified in Fig.\,\ref{figSupMat:DopplerCorrection}, where we show both the uncorrected and corrected resonance frequencies $\omegaResk$. The good agreement of the Doppler-corrected values proves the validity of our approach. All data reported in the main text are Doppler-corrected.
	
	We stress that the value of $\omega_\textrm{D}$ increases with $\kBragg$. As an example, $\omega_\textrm{D}/2\pi$ varies from $15\,$Hz to $40\,$Hz for $\kBragg$ varying from $0.74\,l_z^{-1}$ to $1.74\,l_z^{-1}$ in the measurements of Fig.\,2\,(h). In the analysis of our data, it has thus been important to carefully account for this effect.
	
	\section*{Bragg setup}
	\label{sec:braggsetup-principle}
	\begin{figure*}[t!]
		\includegraphics[width=1\linewidth]{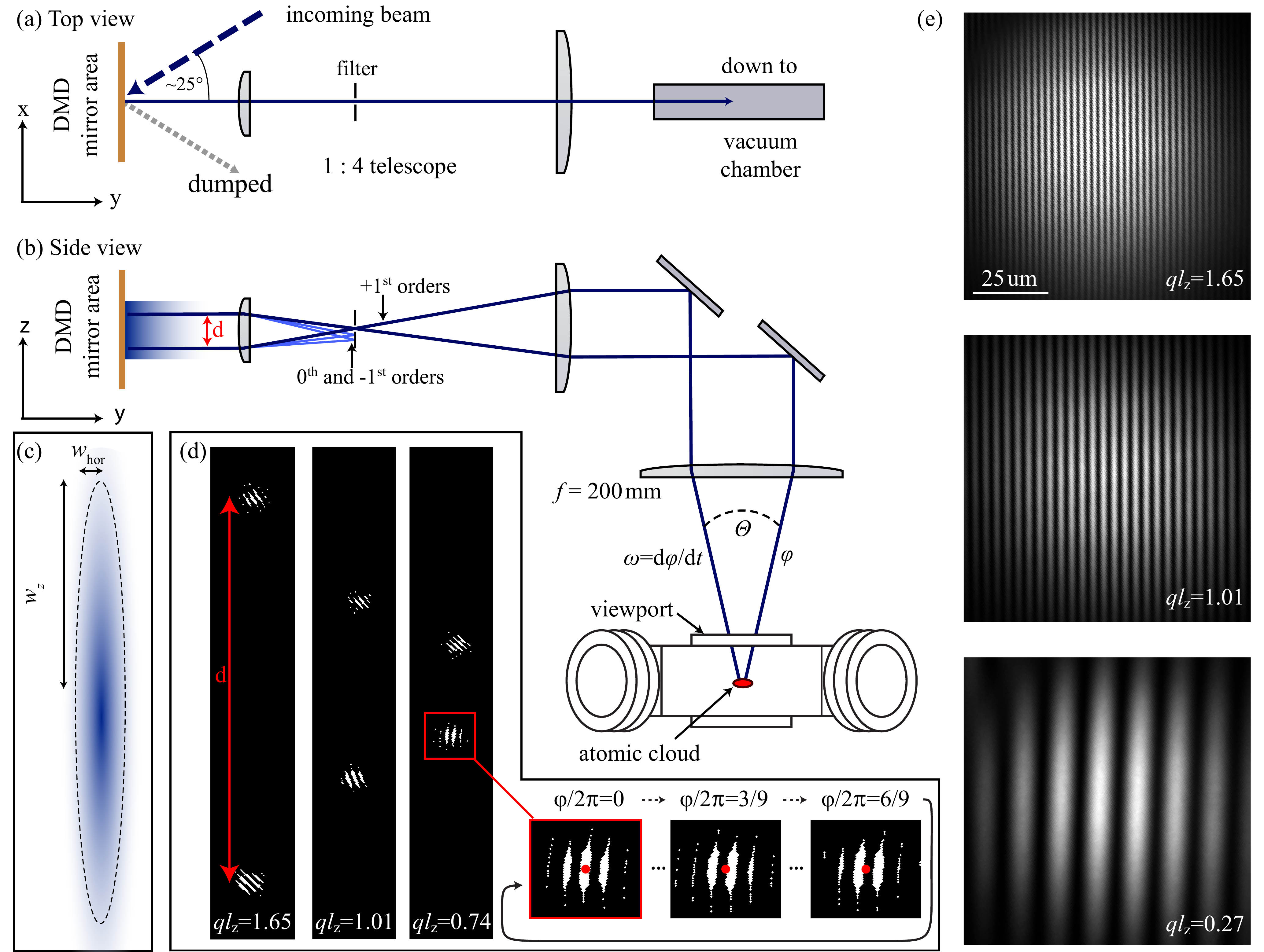}
		\caption{\label{figSupMat:Braggsetup}
			Setup for Bragg spectroscopy: (a) Top view of the Bragg spectroscopy setup, showing the beam path of the incoming beam (dashed arrow), the beams travelling in the Bragg-spectroscopy setup (solid arrow), and the dumped part of the incoming beam (dotted arrow). (b) Side view of the beams travelling in the Bragg spectroscopy setup when a holographic grating pattern is displayed on the DMD (see text for details). (c) Sketch of the elliptic beam shape of the incoming laser beam on the DMD. (d) Examples of binary holograms uploaded on the DMD that allow to create two beams in the Bragg spectroscopy setup, travelling parallel separated by a distance $d$. The three smaller images on the right show a zoom of one part of a hologram with a phase shift of the underlying binary grating, resulting in three distinct phase difference, $\varphi$, of the Bragg beams. (e) Example of light interference patterns at the position of the atomic cloud, obtained from offline calibrations with a CCD camera.
		}
	\end{figure*}
	Our Bragg spectroscopy setup is illustrated in Fig.\,\ref{figSupMat:Braggsetup}\,(a,\,b). It employs a digital micromirror device (DMD), DLP-V9500 from Vialux with $1920\times1080\,$micromirrors. The DMD features a programmable mirror area, consisting of $10.8\times10.8\,$\textmu m-sized micromirrors that can be individually tilted in one of two directions. Depending on the mirror's tilting direction, the incoming light is reflected either into the Bragg spectroscopy setup or on a beam dump. We illuminate the mirror area with a single frequency laser beam of wavelength $\lambdaBragg=401\,$nm. The beam has an elliptic shape with waists of $(w_z,w_\textrm{hor})=(10,1)\,$mm. Here, $w_z$ ($w_\textrm{hor}$) denotes the beam's waist in the $z$ (horizontal) direction; see Fig.\,\ref{figSupMat:Braggsetup}\,(c). The beam is sent on the DMD's mirror area under an angle of $\sim25^{\circ}$ with respect to its perpendicular axis, fulfilling the condition for a blazed grating. This ensures a maximum diffraction efficiency of the incoming beam into the beam path of the Bragg spectroscopy setup.
	
	Following Ref.\,\cite{Zupancic:2016}, the general idea is to use binary holograms that represent maps of titling directions for the micromirrors of the DMD. By placing the DMD in the Fourier-plane of the atoms, the holograms allow for both amplitude and phase modulation of the laser beam at the atoms' position. They consist of an underlying binary grating with two Gaussian envelopes separated by a distance $d$ on the DMD; see Fig.\,\ref{figSupMat:Braggsetup}\,(d). The Gaussian envelopes cut out two beams from the incoming one. Additionally, the envelopes correct for the local intensity inhomogeneities of the incoming beam. After the DMD, the two beams travel parallel with a distance $d$ between each other before being focused in a first optical telescope; see Fig.\,\ref{figSupMat:Braggsetup}\,(b). Due to the binary grating structure of the holograms, each beam splits at the telescope's focus point into a $0^\textrm{th}$ order and $\pm1^\textrm{st}$ side orders. The focus point is used to let only the $+1^\textrm{st}$ order of each beam pass by filtering out the other ones. The two remaining $+1^\textrm{st}$ order beams constitute our two Bragg beams and have a similar Gaussian profile as well as similar intensities. After the telescope they are reflected down to our experimental chamber where a last lens focuses them under an angle $\ThetaBragg$ onto our atomic cloud. At the focus point, matching the atom's position, the beams create an interference pattern with a wavevector $\kBragg$ along $y$. By uploading a hologram with a different $d$ one can change $\ThetaBragg$ and thus $\kBragg$ in an almost continuous manner. We note that due to optical constraints in the experiment (not shown in Fig.\,\ref{figSupMat:Braggsetup}), we can not create interference patterns in the range of $\kBragg=[0.4-0.7]\,l_z^{-1}$.
	
	Furthermore, the relative phase, $\varphi$, of the two beam's wavefronts is directly related to the phase of the applied binary grating on the DMD. It thus allows us to introduce a frequency difference $\omegaBragg=\nicefrac{d\varphi}{dt}$ by displaying a sequence of holograms during the Bragg excitation, where the phase of one of the two Bragg beams is constantly shifted in time; see Fig.\,\ref{figSupMat:Braggsetup}\,(d).
	Practically, we set a sequence of nine holograms, which defines a phase revolution of $2\pi$, and display it in a loop with a fixed rate, $\gamma$. This results in $\omegaBragg=2\pi\gamma/9$. The discrete phase steps of $2\pi/9$ are sufficient to not suffer from higher frequency harmonics in our measurements. We note that $\gamma$ is limited by the maximal refreshing rate of the DMD. Furthermore, the inherent dark time of the DMD at each hologram update results in a decrease of the average intensity of the light grating when increasing $\gamma$. In the experiment, we compensate for this effect by increasing the intensity in the Bragg beams to maintain a constant $\V$.
	
	We further note, that the binary grating in our amplitude holograms take phase aberrations of the optical setup into account and corrects for them~\cite{Zupancic:2016}. These corrections are obtained from offline calibrations with a CCD camera and greatly improve the beam pointing of the individual Bragg beams on the atomic cloud, as all lenses in the optical setup are spherical singlets. In Fig.\,\ref{figSupMat:Braggsetup}\,(e), we show three example images of final interference patterns, obtained with a CCD camera during offline calibrations.
	
	As the Bragg spectroscopy setup uses amplitude holograms, typically only $\sim0.1\%$ of the incoming laser light is used for the Bragg pulse. We take advantage of the strong transition of erbium at 401\,nm. First it allows a wide tuning of $\V$ via the frequency detuning, $\Delta$, of the laser light from the atomic resonance. Second, its short wavelength also leads to a higher maximum $\kBragg$ for a fixed maximum $\ThetaBragg$ as compared to longer laser wavelengths.
	
	In the measurements presented in the main manuscript, the detuning $\Delta =2\pi\times 71(1)\,$GHz	is chosen such that we achieve suitable depths of the Bragg potential (typically $\V \sim h\times10-100\,$Hz), while spontaneous light scattering remains negligible on the experimental time scale. We extract $\V$ via the Kapitza-Dirac-diffraction technique~\cite{Gould1986}. We note that this approach neglects the inhomogeneity of the atomic cloud over the wavelength of the interference pattern and the interactions in the system.
	
	\section*{Image analysis}
	\label{sec:imaging}
	To probe the system's response to the Bragg pulse, we image the atomic cloud after a TOF expansion of 30\,ms. As described in the main text, we perform absorption imaging along the $z$ direction. During the first 15\,ms of the TOF, the $B$-field is kept constant to avoid any sudden change of the dipolar or contact interactions when the atomic density is high. We then set the $B$-field to $B=0.3\,$G and then rotate its direction to obtain a maximal imaging-light scattering cross-section and constant imaging conditions.
	Assuming ballistic expansion, we obtain the mean momentum distribution $n(q_x,q_y)$, by averaging typically four individual images; see Fig.\,1\,(c) in main text. Due to slight variations of the cloud's position from shot to shot, each single image is recentered by extracting the cloud's center from a two-dimensional Gaussian fit. In order to obtain the momentum distribution along the excitation direction of our Bragg pulses, we numerically integrate $n(q_x,q_y)$ along $q_x$ from $[-4.5,+4.5]\,$\textmu m$^{-1}$ and obtain  $n(q_y)$ (1 pixel in our imaging corresponds to $\sim0.32\,$\textmu m$^{-1}$). To extract information on $\DSFBragg$ from $n(q_y)$, we measure either the fraction of excited atoms, $\fexc$, or the momentum variance, $\kySquared$, as introduced in the main text and detailed below.
	
	The procedure used to extract $\kySquared$ depends on $\kBragg$. For $\kBragg l_z>0.7$, we use an asymmetric region of interest (ROI) ranging from $q_y=[-1.9\,$\textmu$\textrm{m}^{-1},\tilde{c}\,\kBragg]$, reflecting the fact that the Bragg excited atoms occur around $q_y=\kBragg>0$. The factor $\tilde{c}$ varies between $[2.5,4.5]$ in order to account for the change in the cloud's momentum width with $\as$ (increasing for increasing $\as$). At low momenta $\kBragg l_z<0.5$, the excited fraction of atoms lies completely within the unscattered peak and only a broadening of the atomic cloud on resonance is observed. Therefore, we choose a symmetric ROI from $q_y=\tilde{c}_\textrm{phon}[-1.6,+1.6]\,$\textmu m$^{-1}$, where $\tilde{c}_\textrm{phon}$ is varied between $[1,3]$ for different $\asf$. We furthermore note that for $\kBragg l_z<0.2$ we are not able to extract a reliable signal in our measurements, as a potential broadening of the atomic cloud on resonance can not be resolved.
	
	In order to extract the excited fraction $\fexc$, we use a bigger ROI that includes the full thermal fraction of atoms and fit a three-Gauss function. The individual Gaussian distributions account for the unscattered atoms in the dBEC $N_0$ (centered at $q_y \approx 0$), the excited atom in the Bragg excitation $\Nexc$ (centered at $q_y \approx \kBragg$) and the broad thermal background. The center positions of the Gaussian distributions for the unscattered atoms and the thermal background are kept the same. The center for fitting the excited fraction is limited to [0.95,\,1.05]\,$\kBragg$. $\fexc$ is then defined as $\fexc=\Nexc/(N_0+\Nexc)$, thus discarding the initial (thermal) population at $\kBragg$ and focusing on the mere fraction of atom promoted during the Bragg pulse.
	
\begin{figure}[t!]
		\includegraphics[width=1\linewidth]{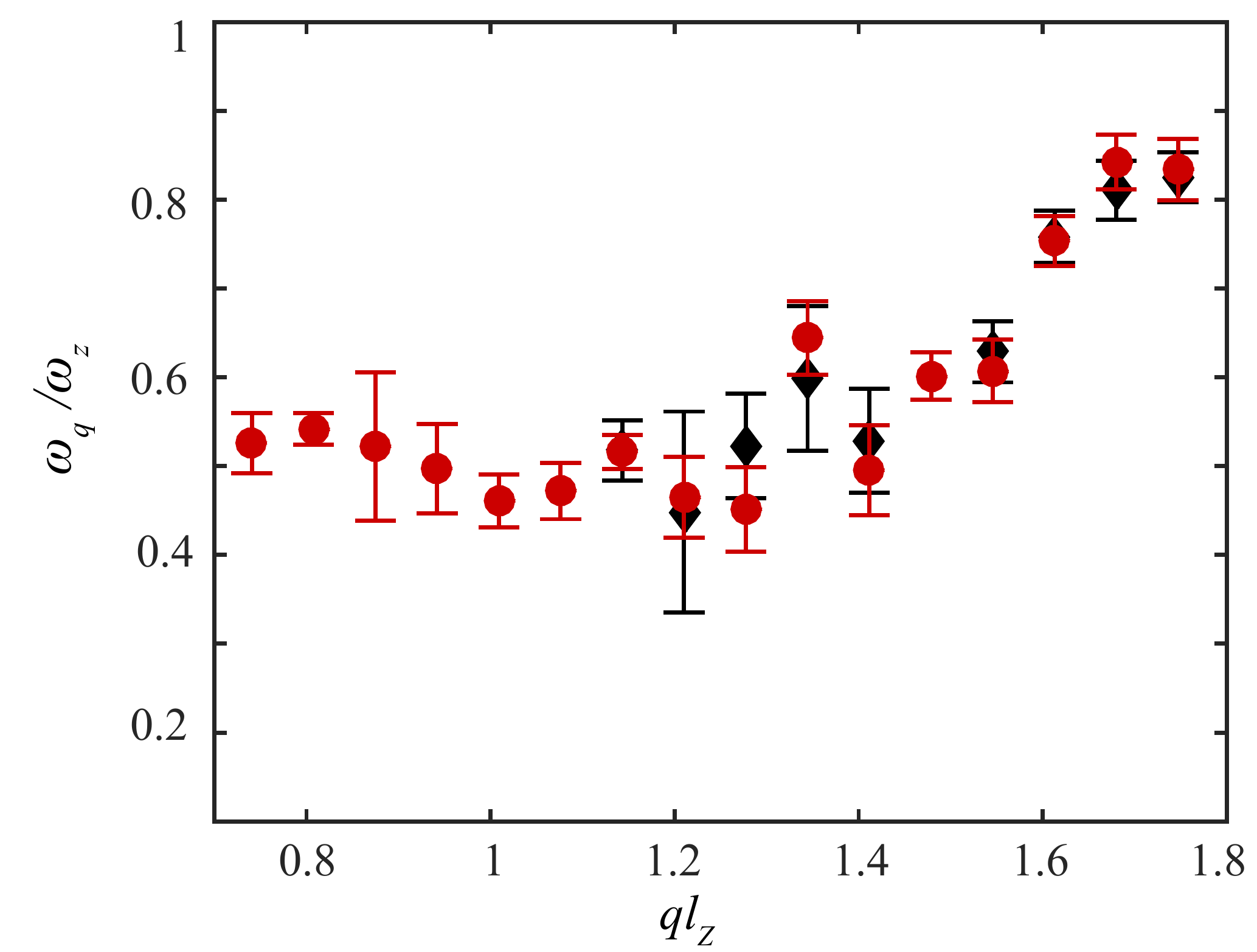}
		\caption{\label{figSupMat:ComparisonRelPopVSKySquared} Resonance frequencies at $\asf=52.5\,a_0$ obtained from $\kySquared$-analysis (cirlces) and $\fexc$-analysis (diamonds). In the latter case, we do not report on values for $\kBragg l_z<1.1$ as $\fexc$ can not be reliably extracted.
		}
	\end{figure}

	We extract $\omegaResk$ and $\fexcres$ by a Gaussian fit to the resonances in $\kySquared$ and $\fexc$ for varying $\omegaBragg$ and fixed $\kBragg$. For too low $\omegaBragg$, the discrete phase steps of our holograms do not provide a well-defined excitation energy over the timescale of the 7-ms Bragg excitation. Hence, in our analysis, we discard points at  $0<\omegaBragg/2\pi\leq 40$\,Hz (corresponding here to the non-Doppler-corrected frequencies; the $0\,$Hz-case is the static case). Nevertheless, we note that an inclusion of these points do not alter the extracted resonance frequencies within their uncertainties. Comparing the $\kySquared$ and $\fexc$ analysis, we also verify that both analysis procedures give the same resonance frequencies within their uncertainties; see Fig.\,\ref{figSupMat:ComparisonRelPopVSKySquared}).

	\section*{Scattering-Length value and its uncertainties}
	\label{sec:uncertantiesScattLength}
	In our experiments we control the contact interaction $\as$ by means of a magnetic Feshbach resonance, centered close to 0\,G~\cite{Chin2010fri}. From previous lattice-spectroscopy measurements, where we probe the excitation gap in the Mott insulator regime~\cite{Chomaz:2016}, we have obtained a precise mapping of the $\as$-to-$B$ conversion. In those measurements, the statistical uncertainty on $\as$ has an average value of $\bar s = 1.8\,a_0$ coming from the uncertainty on the resonance frequency of the Gaussian fit to the spectroscopic data.
	From a fit to the $\as$-data, we obtained a precise $B$-to-$\as$ conversion function for $B$ ranging from 0 to 3\,G ~\cite{Chomaz:2016}. For $\as$ ranging from $80\,a_0$ down to $51\,a_0$ ($B$ from 2.1\,G to 0.21\,G), our conversion function yields a confidence interval of width $\bar c$ varying from $\pm\,0.9$ to $\pm\,1.3\,a_0$. This results into a prediction interval of width $\bar p = \sqrt{\bar c^2 +\bar s^2}$ varying from $\pm\,1.9\,a_0$ to $\pm\,2.1\,a_0$, which defines our conversion uncertainty. In addition we estimate the systematic uncertainty to be $\pm3\,a_0$ with a dominant contribution coming from the uncertainty on the depth of the lattice potential in the spectroscopic measurements (which crucially determine the on-site Wannier function's shape and thus the conversion of the resonance frequency value into $\as$). 
	
	Besides the conversion and systematic uncertainties, statistical uncertainties on $\as$ arise from magnetic field fluctuations and drifts in our experiments. For each dataset we perform independent magnetic-field calibrations by performing radio-frequency (RF) spectroscopy on cold thermal clouds with a 1-ms RF pulse.
	Here we use the same experimental $B$ ramp as for the Bragg measurement and apply a RF pulse of 1\,ms duration, either after holding a time $\tho$ or a time $\tho+\tau$. The mean value of these two measurements is used to extract $\as$. Furthermore, it probes the change of $B$ over the Bragg pulse, which can be up to 3\,mG. Additionally, we have independently estimated that $B$ fluctuates up to $\pm2$\,mG. In total, we consider a $B$ uncertainty of $\pm2.5$\,mG, which we convert in an $\as$ uncertainty based on the $B$-to-$\as$ conversion formula. This can be up to $\pm\,0.2\,a_0$, which is the case for our lowest $\asf$. In conclusion, in the relevant regime for this work, the statistic, conversion and systematic uncertainties on $\as$ are of $\pm\,0.2\,a_0$, $\bar p \sim \pm 2\,a_0$, and $\pm 3.2\,a_0$, respectively.
	
	\section*{Confidence level of the existence of a Roton Minimum}
	\label{sec:statistics}
\begin{figure}[t!]
		\includegraphics[width=1\linewidth]{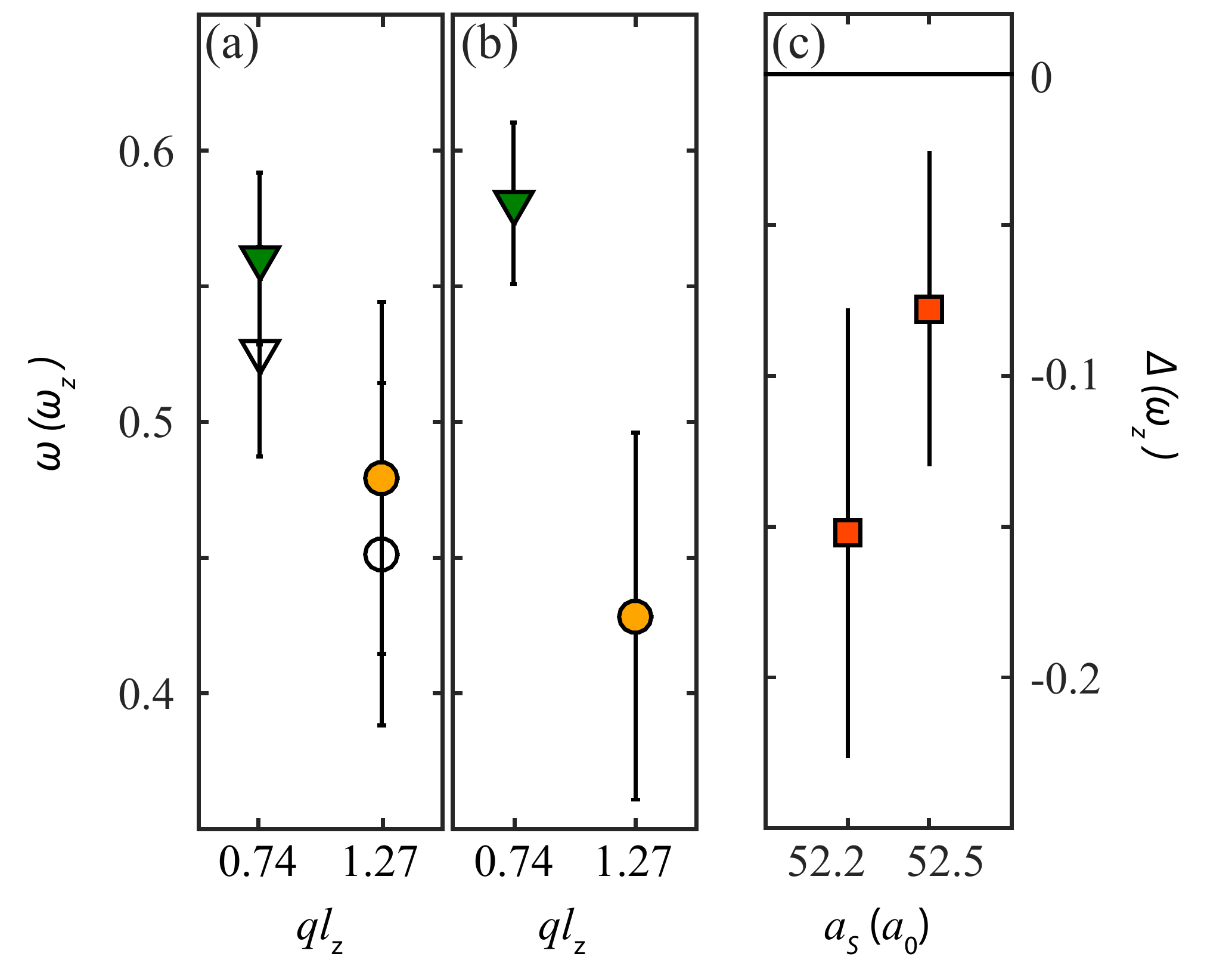}
		\caption{\label{figSupMat:statAnalysis}Analysis on the extracted resonance frequencies $\omegaResMax$ (triangles) and $\omegaResRot$ (circles) for $52.5(2)\,a_0$ (a) and $52.2(2)\,a_0$ (b). In (a, b) filled symbols show data from Fig.\,3, where as  empty symbols show data points from  Fig.\,2~(h). Error bars represent $\pm$ one standard deviation of the corresponding Gaussian fits. (c) shows the corresponding differences $\Delta$ (after averaging for $52.5\,a_0$) together with its uncertainties, deduced from standard error propagation.
		}
	\end{figure}

	In order to confirm the existence of a local minimum in the excitation spectrum of our dBECs, we compare the maxon with the roton energy and extract a confidence level from a statistical analysis. First, we focus on a scattering length value of $\as =52.5(2)\,a_0$ for which two sets of data are available (from Fig.\,2\,(h) and inset of Fig.\,3). From both resonance frequencies $\omegaResMax$ and $\omegaResRot$, we obtain the corresponding mean values $\tilde{\omega}_\textrm{m}$ and $\tilde{\omega}_\textrm{rot}$. Calculating the difference
	$\Delta=\tilde{\omega}_\textrm{m}-\tilde{\omega}_\textrm{rot}=-0.08(5)\omega_z$ reveals the existence of a roton minimum with a 93\% confidence level; see Fig.\,\ref{figSupMat:statAnalysis}\,(a,c).
	The existence of a roton minimum in the spectrum is even more evident by analyzing $\Delta=\omegaResMax-\omegaResRot$ at $52.2(2)\,a_0$, where the minimum is deeper. Here, we find $\Delta=-0.15(7)\omega_z$ giving a confidence level of 98\% for the existence of a minimun in the spectrum of a stable dBEC; see Fig.\,\ref{figSupMat:statAnalysis}\,(b,c).

	\section*{Theory}
	\label{sec:theory}
	To compare our experiment with theory predictions, we perform numerical calculations of the dynamic structure factor following the procedure detailed in the supplementary information of Ref.\,\cite{Chomaz:2018}. The calculations are based on a Bogoliubov treatment of an extended GPE with energy functional $\hat{H}_{\mathrm{GP}}[\psi]$, for which our \Er dBEC is the ground-state. The classical field $\psi$ describes the macroscopic wavefunction of $N$ atoms and is normalized to $N$. The dBEC state, $|0\rangle$, corresponds to the wavefunction $\psi_0$. The Bogoliubov analysis gives access, in second-order perturbation, to the discrete modes, $|l\rangle$, of the dBEC's excitation spectrum, to their energies $\hbar\omega_l$ and the Bogoliubov spatial amplitudes $u_l$ and $v_l$.  
	
	In the present calculations, we can in principle account for the effect of the quantum fluctuations, by including the Lee-Huang-Yang term $\Delta \mu[n] = 32 g (n \as)^{3/2}(1 + 3 \edd^2 / 2) / 3 \sqrt{\pi}$ in $\hat{H}_{\mathrm{GP}}[\psi]$ (here $g = \frac{4\pi \hbar^2 \as}{m}$ and $n = |\psi|^2$). $\Delta \mu[n]$ has been obtained under a local density approximation~\cite{Pelster:2011,Pelster:2012} when computing the Bogoliubov modes. The relevance of the inclusion of such a potential correction has been demonstrated in various studies of dipolar Bose gases close to the mean field instability~\cite{Waechtler:2016,Waechtler:2016b,Bisset:2016, Chomaz:2016, Schmitt:2016, Chomaz:2018}. However, as described in the main text, a better agreement with experimental data, close to the instability, is found instead by omitting the Lee-Huang-Yang term. $\Delta \mu[n]$ is then only included in $\hat{H}_{\mathrm{GP}}$ for computing the dotted line and the corresponding yellow shading in Fig.\,3 of the main text. It is discarded from all other theory calculations reported in the main text.
	
	The knowledge of the Bogoliubov modes allows one to compute the bare zero-temperature dynamic structure factor $\DSFbrut$, which is defined as~\cite{Brunello:2001, Blakie:2002},
	\begin{equation}\label{EqZeroTStrucFacdef}
		\DSFbrut = \sum_l \left|\langle l| \delta {\hat n}_{\bs q}^\dagger |0\rangle \right|^2 \delta(\omega - \omega_l),
	\end{equation} 
	with $\delta {\hat n}_{\bs q}$ being the density fluctuation operator in momentum space:
	\begin{eqnarray}\label{nq}
		\delta {\hat n}_{\bs q} = \int \mathrm{d}\rvec \ \mathrm{e}^{\mathrm{i} \vq\cdot \rvec}\left({\hat \psi}^\dagger(\rvec){\hat \psi}(\rvec)-\langle0|{\hat \psi}^\dagger(\rvec){\hat \psi}(\rvec)|0\rangle\right),
	\end{eqnarray}
	and ${\hat \psi}$ is the field operator. The matrix elements of the density fluctuation operator are computed as
	\begin{equation}\label{EqStrucFac}
		\langle l| \delta {\hat n}_{\vq}^\dagger |0\rangle =\int \mathrm{d}\rvec [u_l^*(\rvec) +v_l^*(\rvec)]\mathrm{e}^{\mathrm{i} \vq \cdot \rvec}\psi_0(\rvec).
	\end{equation}
	Considering Eq.\,\eqref{EqZeroTStrucFacdef}, one sees that $\DSFbrut$ consists of infinitely narrow peaks centered around $\omega = \omega_l$. The integrated amplitude of each peak corresponds to the contribution of the mode $l$ to the quantum density fluctuations of the dBEC at momentum $q$. 
	
	For our experimental probing, the relevant quantity is the Fourier-broadened structure factor, $\DSF=\Big[\tau\textrm{sinc}^2(\tauBragg\omega'/2)*S_0(\vq,\,\omega')\Big](\omegaBragg)$, where $*$ denotes a convolution over $\omega'$. This ultimately writes
	\begin{equation}\label{EqStrucFacdef}
		\DSF = \sum_l \left|\langle l| \delta {\hat n}_{\bs q}^\dagger |0\rangle \right|^2 \tau\textrm{sinc}^2(\tauBragg(\omegaBragg-\omega_l)/2).
	\end{equation} 
	$\DSF$ shows the same peaks in frequency as $\DSFbrut$ but broadened with a typical width $1/\tau$. The amplitude of $\DSF$ on resonance  matches the contribution of the mode $|l\rangle$ to the quantum density fluctuations of the dBEC at momentum $q$, multiplied by the Bragg pulse duration, $\tau$.
	
	\section*{Connection between measured quantities and DSF}
	
	In our experiment, we probe the dynamic structure factor either via the fraction of atoms excited from the dBEC peak at $q_y=0$ to the Bragg peak at $q_y=q$ or via the momentum variance along the $y$ axis. Following Refs.\,\cite{Brunello:2001, Blakie:2002}, we derive the relations of our observables to $\DSF$. The occupation of each mode after the pulse is given by (Ref.\,\cite{Blakie:2002}, Eq.\,(2.31)):
	\begin{eqnarray}
		\fexc_l &=&  \frac{\langle N_l(t=\tau)\rangle -\langle N_l(t=0) \rangle}{N}\label{fexcl1}\\
		&=&\frac{\pi^2 \V^2\tauBragg}{h^2} \left|\langle l| \delta {\hat n}_{\bs q}^\dagger |0\rangle \right|^2 \tau\textrm{sinc}^2(\tauBragg(\omegaBragg-\omega_l)/2), \label{fexcl}
	\end{eqnarray}
	with $\langle N_l(t)\rangle$ being the mean number of atoms in the mode $l$ at time $t$ ($t=0$ matching the beginning of the Bragg pulse). Equation~(1) of the main text is then found by simply summing $\fexc=\sum_l \fexc_l$ and using Eq.\,\eqref{EqStrucFacdef}.
	We note that in our data analysis, the thermal (i.\,e.\,initial) population at $q_y=\kBragg$ is encompassed in the broad background Gaussian and thus excluded from the definition of $\fexc$, similarly to Eq.\,\eqref{fexcl1}.
	
	For the momentum variance, the situation is more complex. Assuming a fully ballistic expansion and a linear perturbation regime, $\hbar^2\kySquared/2m$ matches the energy transferred during the Bragg pulse. For each mode, the energy transferred during the pulse writes  $\fexc_l N \hbar \omega_l$. Using Eq.\,\eqref{fexcl}, one finds
	\begin{multline}\label{eq:kysquared}
		\qquad\kySquared  - \kySquared_0 = \\ \frac{4\pi^3 m \V^2\tauBragg N }{h^3}\sum_l \left|\langle l| \delta {\hat n}_{\bs q}^\dagger |0\rangle \right|^2 \tau\omega_l\textrm{sinc}^2(\tauBragg(\omegaBragg-\omega_l)/2).
	\end{multline}
	where $\kySquared_0$ is the value of $\kySquared$ for the dBEC (of $N$ atoms) in absence of a Bragg pulse (at $t=0$) and typically depends on the value of $\as$. Because of the multiplication of $\omega_l$ in the sum, $\kySquared$ can only be related to $\DSF$ approximately. When only one mode contributes significantly to $\DSF$, at a given $(\bs{\kBragg},\omegaBragg)$ one can write 
	\begin{equation}\label{eq:kysquared}
		\kySquared  - \kySquared_0 \approx  \frac{8\pi^4 m \V^2\tauBragg N \Eexc(\vq)}{h^4}\DSF.
	\end{equation}
	Note that, for a fixed $\bs\kBragg$, $\Eexc(\vq)$ is a constant multiplying the overall amplitude but not affecting the peak position in $\omegaBragg$. We highlight again that, in the experiment, we observe no significant difference in the extracted $\Eexc(\kBragg)$ when analyzing $\fexc$ or $\kySquared$; see Fig.\,\ref{figSupMat:ComparisonRelPopVSKySquared}.
		
\end{document}